\documentclass[11pt]{article}
\usepackage{geometry}
\usepackage{here}
\usepackage{graphicx}
\usepackage{amsmath,amssymb}
\usepackage{cite}
\usepackage{bm}
\newcommand{\lw}[1]{\smash{\lower 1.5ex\hbox{#1}}}

\begin{document}

\title{
\vspace{-23mm}
Two-component model description of Bose-Einstein correlations in pp collisions at 13 TeV measured by the CMS Collaboration at the LHC}

\author{Takuya Mizoguchi$^{1}$, Seiji Matsumoto$^{2}$, and Minoru Biyajima$^{3}$\\
{\small $^{1}$National Institute of Technology, Toba College, Toba 517-8501, Japan}\\
{\small $^{2}$School of General Education, Shinshu University, Matsumoto 390-8621, Japan}\\
{\small $^{3}$Department of Physics, Shinshu University, Matsumoto 390-8621, Japan}}

\date{}
\maketitle

\begin{abstract}
Using the two-component model, we analyze Bose-Einstein correlations in pp collisions at the center-of-mass energy of 13 TeV, measured by the CMS Collaboration at the LHC, and compare results with the $\tau$-model. We utilize data described by the double ratios with an average pair transverse momentum $0\le k_T\le 1.0$ GeV and six intervals described by the reconstructed charged particle multiplicity as $N_{\rm trk}^{\rm offline}$. The estimated ranges are 1-4 fm for the magnitude of extension of emitting source expressed by the exponential function $\exp(-RQ)$ and 0.4-0.5 fm for that by the Gaussian distribution $\exp(-(RQ)^2))$, respectively. Moreover, we estimate the upper limits of the 3-pion BEC to test the two-component model and investigate the role of the long-range correlation.
\end{abstract}

%SECTION1-----------------------------------------------------------------------
\section{\label{sec1}Introduction}
\begin{figure}[H]
  \centering
  \includegraphics[width=0.48\columnwidth]{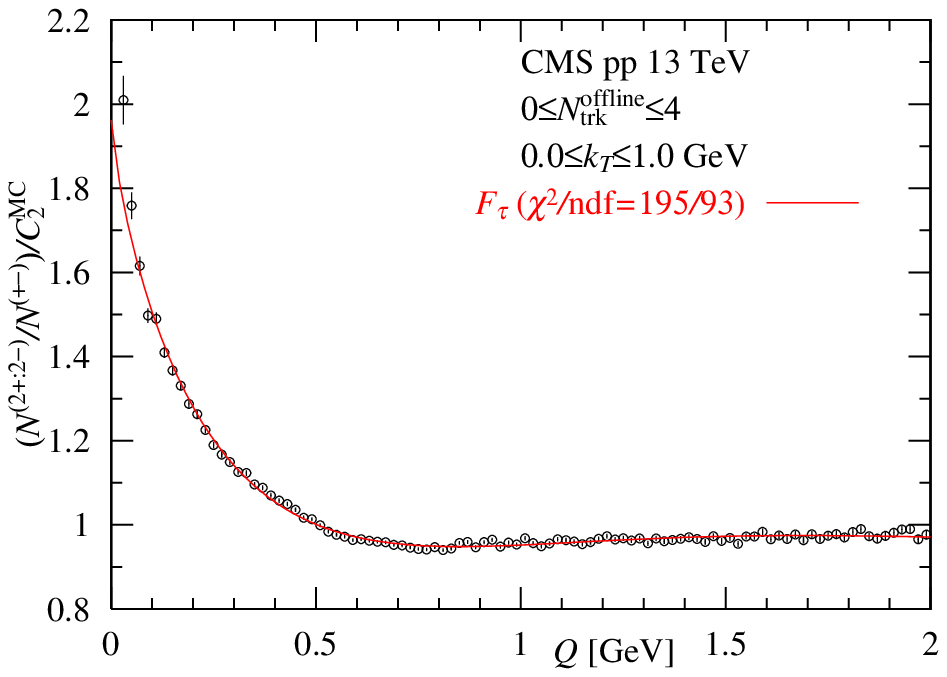}
  \includegraphics[width=0.48\columnwidth]{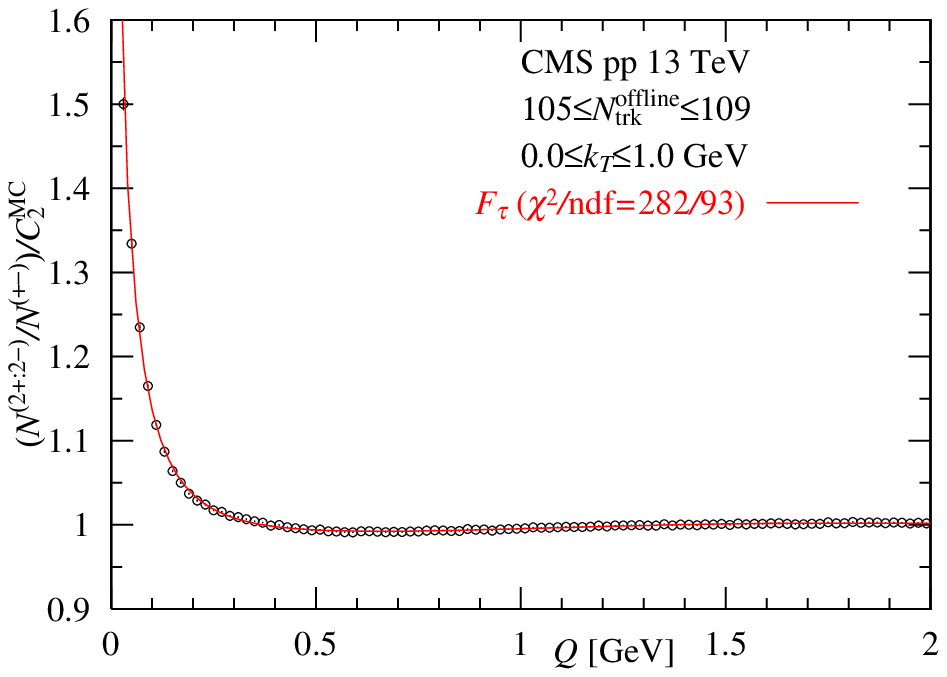}
  \caption{\label{fig1}Fit to the BEC measurements by CMS in pp collisions at 13 TeV by Eq.~(\ref{eq1}). $C_2^{MC}\equiv N_{\rm MC}^{(2+:2-)}/N_{\rm MC}^{(+-)}$, where $N_{\rm MC}$'s mean the numbers of the same charged and opposite charged pairs recorded in MC simulations.}
\end{figure}

This article investigates the Bose-Einstein correlations (BEC) described by double ratios (DRs) in pp collisions at the center-of-mass energy 13 TeV, obtained by the CMS Collaboration at the LHC~\cite{CMS:2019fur}. The DR is defined by two single rations (SR's), i.e., ${\rm C_2^{data}}=N^{(2+:2-)}/N^{\rm (+-)}$ and ${\rm C_2^{MC}}=N_{\rm MC}^{(2+;2-)}/N_{\rm MC}^{(+-)}$, where $N$'s mean the number of events in data and the Monte Carlo simulation. The suffixes ${\rm (2+:2-)}$ and ${\rm (+-)}$ mean the charge combinations. Therein, CMS Collaboration only reports $\chi^2/$ndf (number of degrees of freedom) values obtained using the $\tau$-model. Here, we analyze the DRs at an average pair transverse momentum $0\le k_T\le 1.0$ GeV ($k_T=|{\bf p}_{T,1}+{\bf p}_{T,2}|/2$), and six intervals expressed by means of constraint $a\le N_{\rm trk}^{\rm offline}\le b$ as illustrated in Fig.~\ref{fig1}. The formula used in the CMS analysis~\cite{CMS:2011nlc} is 
\begin{eqnarray}
F_{\tau}=C[1+\lambda\cos( (r_0 Q)^2+\tan(\alpha_{\tau}\pi/4)(Qr)^{\alpha_{\tau}})e^{-(Qr)^{\alpha_{\tau}}}]\cdot (1+\delta Q)
\label{eq1}
\end{eqnarray}
where $\lambda$, $r_0$, $r$, and $\alpha_{\tau}$ are parameters introduced in the stable distribution based on stochastic theory, namely the degree of coherence, two interaction ranges, and the characteristic index, respectively (see, also Refs.~\cite{Takayasu:1990aa,Ito:2004aa}). $Q=\sqrt{-(p_1-p_2)^2}$ is the magnitude of the 4-momentum transfer between two pions.  The last term $(1+\delta Q)$ is named the long range correlation with the index (linear) (LRC$_{\rm (linear)}$). Our estimated values are presented in Table~\ref{tab1}.

%FIG1---------------------------------------------------------------------------

%TAB1---------------------------------------------------------------------------
\begin{table}[H]
\centering
\caption{\label{tab1}Fit parameters to the CMS BEC measurements in pp collisions at 13 TeV at $0.0\le k_T\le 1.0$ GeV by Eq.~(\ref{eq1}). $\delta$ values are estimated (top to bottom): $-0.016\pm 0.004$, $0.03\pm 0.01$, $0.005\pm 0.005$, $(-1.2\pm 0.1)\times 10^{-3}$, $(1.3\pm 0.2)\times 10^{-3}$, and $0.002\pm 0.001$.} 
\vspace{2mm}
\renewcommand{\arraystretch}{1.2}
\begin{tabular}{cccccccc}
\hline
$N_{\rm trk}^{\rm offline}$
& $r_0$ (fm)
& $r$ (fm) 
& $\lambda$
%& $C$    
& $\alpha_{\tau}$
%& $\delta$
& $\chi^2$/ndf
& $\chi^2$(CMS)\\
\hline
  $ 0-4$
& $ 0.139\pm 0.021$
& $ 0.93\pm 0.06$
& $ 0.96\pm 0.05$
%& $ 1.00\pm 0.00$
& $ 0.781\pm 0.026$
%& $-0.02\pm 0.00$
&   195/93
&   195\\

  $ 10-12$
& $ 0.244\pm 0.004$
& $ 9.08\pm 1.40$
& $ 2.43\pm 0.23$
%& $ 0.97\pm 0.00$
& $ 0.420\pm 0.013$
%& $ 0.03\pm 0.00$
&   140/93
&   140\\

  $ 31-33$
& $ 0.232\pm 0.005$
& $ 21.8\pm 4.0$
& $ 3.36\pm 0.37$
%& $ 0.99\pm 0.00$
& $ 0.377\pm 0.011$
%& $ 0.01\pm 0.00$
&   135/93
&   135\\

  $ 80-84$
& $ 0.224\pm 0.001$
& $ 43.7\pm 2.5$
& $ 4.48\pm 0.15$
%& $ 1.00\pm 0.00$
& $ 0.351\pm 0.003$
%& $ 0.00\pm 0.00$
&   899/93
&   902\\

  $ 105-109$
& $ 0.216\pm 0.003$
& $ 47.0\pm 5.2$
& $ 4.71\pm 0.31$
%& $ 1.00\pm 0.00$
& $ 0.352\pm 0.005$
%& $ 0.00\pm 0.00$
&   282/93
&   281\\

  $ 130-250$
& $ 0.228\pm 0.013$
& $ 53.3\pm 19.9$
& $ 5.32\pm 1.27 $
%& $ 1.00\pm 0.00 $
& $ 0.353\pm 0.020$
%& $ 0.00\pm 0.00 $
&   84.5/93
&   84\\
\hline
\end{tabular}
\end{table}
Because estimated values of all parameters by $\tau$-model, i.e., Eq.~(\ref{eq1}), have not been presented in Ref.~\cite{CMS:2019fur}, it is difficult to draw physical picture through the analyses of BEC in pp collisions at 13 TeV. Thus for this aim, we present them in Table~\ref{tab1}. Table~\ref{tab1} shows that the $\chi^2/$ndf values obtained from our analysis are consistent with those reported by the CMS Collaboration~\cite{CMS:2019fur}. In other words, through concrete figures in Table~\ref{tab1}, we are able to consider physical picture based on the $\tau$-model.

As indicated in Table~\ref{tab1}, the interaction ranges of the Levy-type form ($e^{-(Qr)^{\alpha}}$) increase as the interval containing $N_{\rm trk}^{\rm offline}$ increases. The estimated values $r=20\sim 50$ fm appear large for pp collisions at 13 TeV.

This paper also investigates this issue from a different perspective, focusing on the collision mechanism. Three processes occur in collisions at the LHC~\cite{Navin:2011oct,Zborovsky:2013tla,Biyajima:2019wcb}: the non-diffractive dissociation, the single-diffractive dissociation, and the double-diffractive dissociation (DD). BEC are related to the chaotic components of particle production. Since the contribution from the DD is Poissonian~\cite{Biyajima:2019wcb}, there is no effect to the BEC. Thus we calculated the following two-component model correlation function~\cite{Biyajima:2019wcb,Mizoguchi:2019cra} (see also empirical Refs.~\cite{AxialFieldSpectrometer:1987otj,EHSNA22:1993ugp,Lorstad:1988ff}),
\begin{eqnarray}
{\rm CF_{II}}=1+\lambda_1E_{\rm BE_1}+\lambda_2E_{\rm BE_2}.
\label{eq2}
\end{eqnarray}
The exchange function is the Fourier transform of the space-time region emitting bosons (mainly pion) with overlapping wave functions. For the exchange functions $E_{\rm BE_1}$ and $E_{\rm BE_2}$, we assign the following two functions~\cite{Shimoda:1992gb},
\begin{eqnarray}
\exp(-R_1Q)\quad{\rm and}\quad \exp\left(-(R_2Q)^2\right)
\label{eq3}
\end{eqnarray}
characterizing the exponential and Gaussian type of BEC. Thus $R_1$ and $R_2$ mean the extensions of the sources~\cite{Shimoda:1992gb}. Concerning with two kinds of exchange functions, see also the different approach\cite{Khoze:2016hns}.

Moreover, we discuss the LRC's below. Three decades ago, the OPAL Collaboration~\cite{Acton:1991xb} adopted LRC$_{(\rm OPAL)}=c(1+\delta Q+\varepsilon Q^2)$ to improve the linear form LRC$_{(\rm linear)}=C(1+\delta Q)$. Recently we proposed the inverse power series form LRC$_{(\rm p.s.)}=C/[1-\alpha Q\exp(\beta Q)]$~\cite{Mizoguchi:2021slz}, because the number of parameters ($\alpha$ and $\beta$) is the same as the LRC$_{(\rm OPAL)}$'s, and it converges to $C$ as $Q$ is large. Taking into account of those investigations and mathematical descriptions shown in Ref.~\cite{CMS:2019fur}, i.e., the distribution of opposite charged pion pair $N^{(+-)} = C[1+a\exp{(-bQ^2)}]$ and so on, we propose the following form:
\begin{eqnarray}
{\rm LRC_{(Gauss)}}=\frac{C}{1.0+\alpha\exp(-\beta Q^2)}.
\label{eq4}
\end{eqnarray}
This function converges to $C$ as $Q$ is large and behaves as $C[1-\alpha(1-\beta Q^2)+\cdots]$, $Q$ being small. In Table \ref{tab1A}, we compare our approach with formulas shown in Ref.~\cite{CMS:2019fur}.

%TAB1A---------------------------------------------------------------------------
\begin{table}[H]
\centering
\caption{\label{tab1A}Comparison of our approach with formulas utilized by CMS Collaboration~\cite{CMS:2019fur}.} 
\vspace{2mm}
\renewcommand{\arraystretch}{1.2}
\begin{tabular}{c|l|l}
\hline

& formulas
& cf.\\
\hline
& \quad ${\rm CF_{II}\times LRC}$
& 1) ${\rm CF_{II}}$ is reflecting to three kinds of multiplicity\\

Our
& where
& distributions of the ND, SD, and DD in pp collisions.\\

$\!\!\!$approach
& \quad ${\rm LRC_{(Exp)}}=\dfrac 1{1+\alpha e^{-\beta Q}}$,
& 2) Through the generalization of ${\rm LRC_{(OPAL)}}$\\

& or
& $=1+\delta Q + \varepsilon Q^2$ in e$^+$e$^-$ annihilation at Z$^0$-pole~\cite{Acton:1991xb},\\

& \quad ${\rm LRC_{(Gauss)}}=\dfrac 1{1+\alpha e^{-\beta Q^2}}$.
& we obtained ${\rm LRC_{(Exp)}}$~\cite{Mizoguchi:2021slz}.\\

&
& 3) Referring to mathematical descriptions on $F_{2N}$ and \\

& 
& $F_{2D}$ in Ref~\cite{CMS:2019fur}, ${\rm LRC_{(Gauss)}}$ is proposed for pp collisions.\\

\hline

& 1) Distributions $N^{(2+:2-)}$, $N^{(+-)}$, 
& 1) ${\rm CF_{I}}=1+\lambda E_{\rm BE}$, where $E_{\rm BE}=\exp(-RQ)$~\cite{CMS:2019fur}. \\

CMS
& $N_{\rm MC}^{(2+:2-)}$ and $N_{\rm MC}^{(+-)}$ are assumed 
& 2) Provided that $N_{\rm MC}^{(+-)}\cong N^{(+-)}$ and the cross term \\

& as follows: ${\rm CF_I}\cdot C(1+ae^{-bQ^2})$, 
& ($\lambda aE_{\rm BE}\cdot e^{-bQ^2}$) is small, we obtain\\

& $C'(1+a'e^{-b'Q^2})$, $C_M(1+a_Me^{-b_MQ^2})$ 
& $\dfrac{F_{2N}}{F_{2D}}\cong \dfrac{{\rm CF_I}\times (1+ae^{-bQ^2})}{1+a_Me^{-b_MQ^2}}\cong {\rm CF_{II}\times LRC_{(Gauss)}}$.\\

& and $C_M'(1+a_M'e^{-b_M'Q^2})$, respectively.
& Thus, $\alpha$ and $\beta$ in Eq.~(\ref{eq4}) are approximately identified \\

& 2) SR's $F_{2N}=N^{(2+:2-)}/N^{(+-)}$ and 
& with $a_M$ and $b_M$ describing the Monte Carlo events \\

& $F_{2D}=N_{\rm MC}^{(2+:2-)}/N_{\rm MC}^{(+-)}$ are used for 
& in $F_{2D}$, respectively.\\

& analysis of data of DR by the ratio 
& 3) Their Monte Carlo events are calculated by \\

& $F_{2N}/F_{2D}$. 
& PYTHIA6. See Fig.~3 in Ref.~\cite{CMS:2011nlc}.\\

& 3) $\tau$-model is also used for data of DR.
& \\

\hline
\end{tabular}
\end{table}

In the second section, we analyze the BEC at 13 TeV using Eqs.~(\ref{eq2})--(\ref{eq4}). In the third section, we present our predictions for 3-pion BEC using the two-component model. In the final section, we provide concluding remarks. Appendix~\ref{secA} presents an analysis of BEC at 13 TeV using the $\tau$-model with Eq.~(\ref{eq4}). In Appendix~\ref{secB}, we reanalyze the CMS BEC at 0.9 and 7 TeV utilizing Eq.~(\ref{eq4}), because in previous works~\cite{Biyajima:2019wcb,Mizoguchi:2019cra}, we used LRC$_{(\rm linear)}=C(1+\delta Q)$.

%SECTION2-----------------------------------------------------------------------
\section{\label{sec2}Analysis of BEC at 13 TeV using Eqs.~(\ref{eq2})--(\ref{eq4}).}
\begin{figure}[H]
  \centering
  \includegraphics[width=0.48\columnwidth]{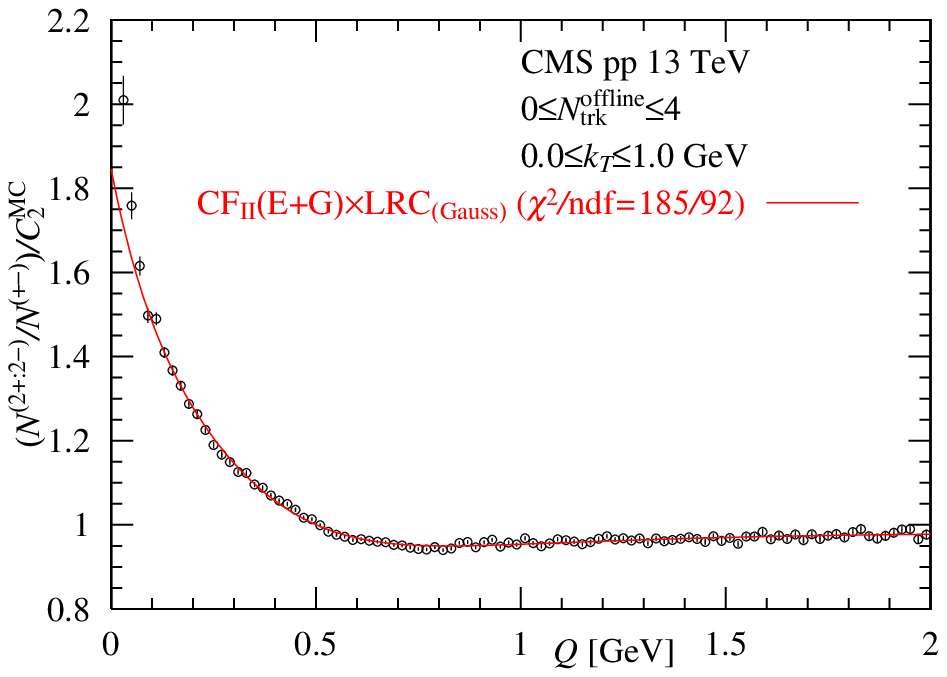}
  \includegraphics[width=0.48\columnwidth]{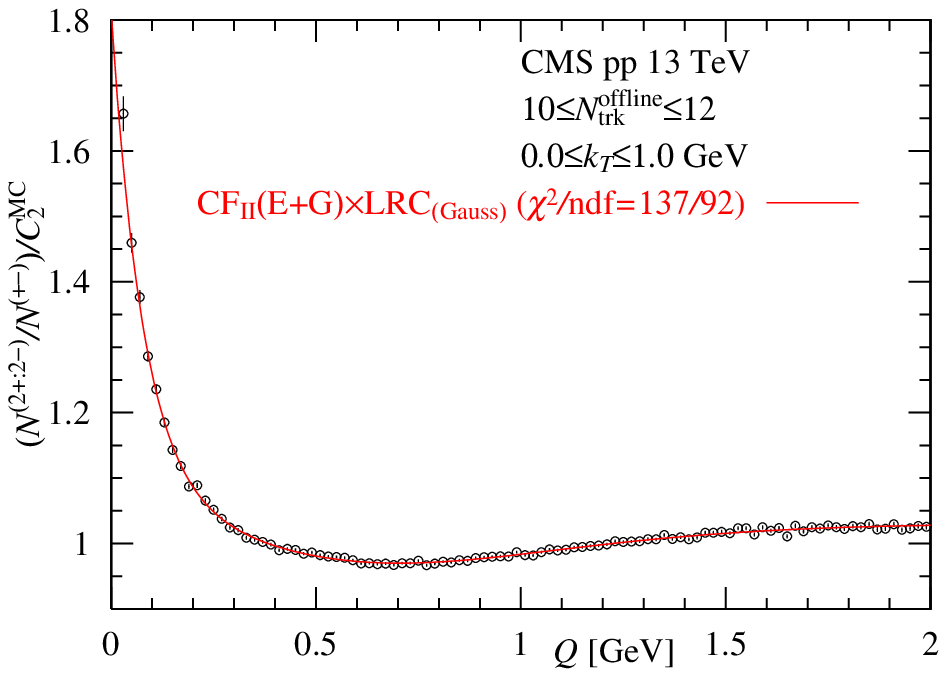}\\
  \includegraphics[width=0.48\columnwidth]{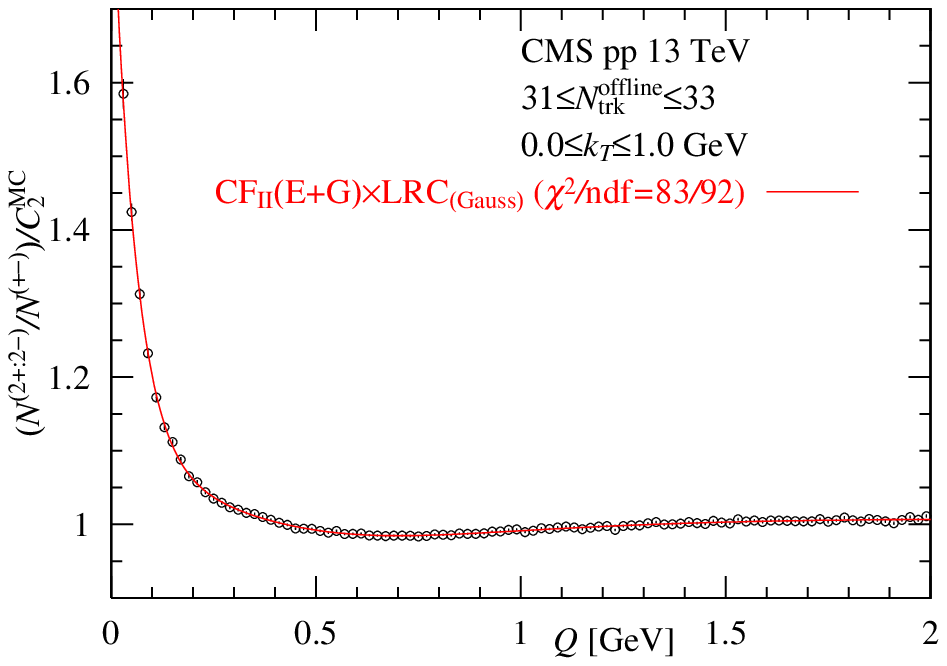}
  \includegraphics[width=0.48\columnwidth]{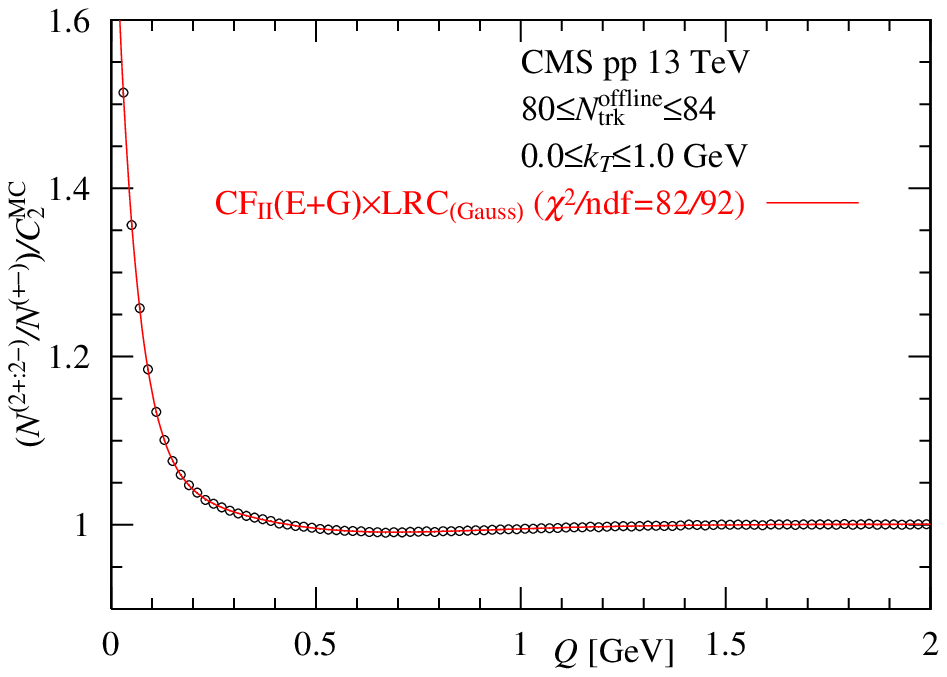}\\
  \includegraphics[width=0.48\columnwidth]{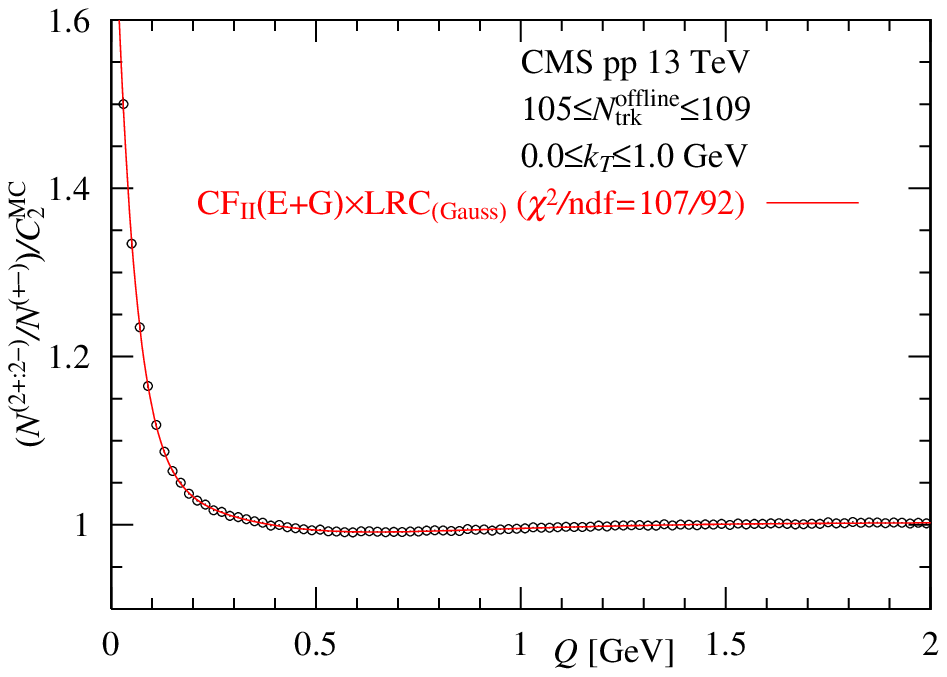}
  \includegraphics[width=0.48\columnwidth]{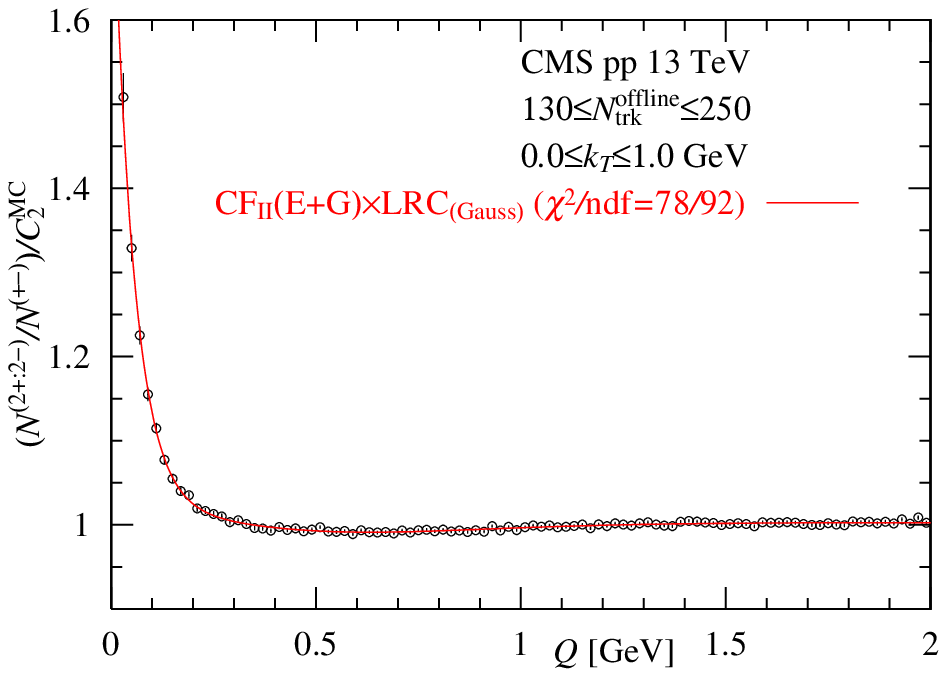}\\
  \caption{\label{fig2}Fit to the BEC measurements by CMS in pp collisions at 13 TeV by Eqs.~(\ref{eq2})--(\ref{eq4}).}
\end{figure}

\begin{figure}[H]
  \centering
  \includegraphics[width=0.6\columnwidth]{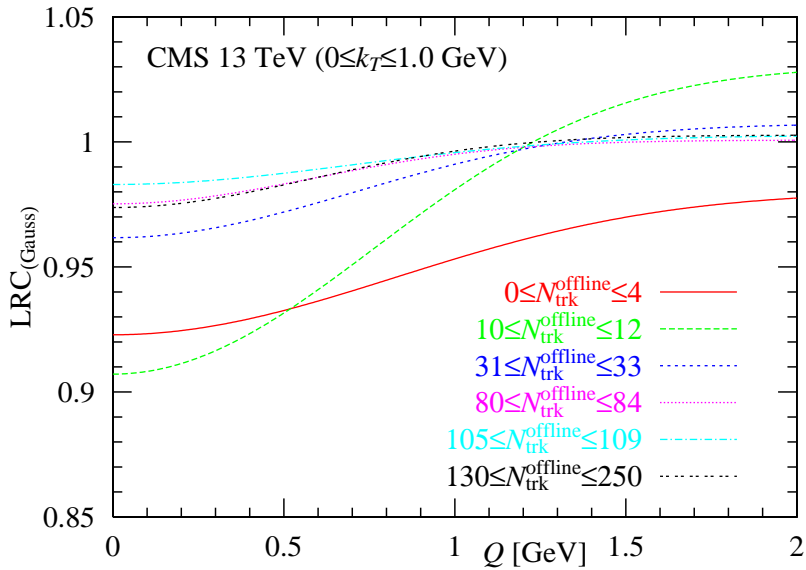}
  \caption{\label{fig3}The long-range correlations (LRCs), see Eq.~(\ref{eq4}) for six intervals.}
\end{figure}

Considering the results of the CMS BEC at 7 TeV in Ref.~\cite{Biyajima:2019wcb}, we assume a combination of exponential function and Gaussian distribution, as this combination has shown the valuable role. Moreover, it is worthwhile to mention that Shimoda et al. in Ref.~\cite{Shimoda:1992gb} investigated several possible distributions for $E_{\rm BE}$'s. Our results are presented in Fig.~\ref{fig2} and Table~\ref{tab2}. We observe extraordinary behaviors in the two intervals, $0\le N_{\rm trk}^{\rm offline}\le 4$ and $10\le N_{\rm trk}^{\rm offline}\le 12$, of the LRC shown in Fig.~\ref{fig3}.

As indicated by Fig.~\ref{fig2} and Table~\ref{tab2}, the two-component model with Eqs.~(\ref{eq2})--(\ref{eq4}) effectively characterizes three intervals: $31\le N_{\rm trk}^{\rm offline}\le 33$, $80\le N_{\rm trk}^{\rm offline}\le 84$, and $105\le N_{\rm trk}^{\rm offline}\le 109$. 

%FIG2---------------------------------------------------------------------------

%TAB2---------------------------------------------------------------------------
\begin{table}[H]
\centering
\caption{\label{tab2}Fit parameters of the CMS measurements of BEC in pp collisions at 13 TeV ($0.0\le k_T\le 1.0$ GeV) by Eqs.~(\ref{eq2})--(\ref{eq4}). Three constraints are used: $\lambda_1\le 1.0$, $\lambda_2\le 1.0$, and $\lambda_1+\lambda_2\le 1.0$. The $p$-values for the three intervals $31\le N_{\rm trk}^{\rm offline}\le 33$, $80\le N_{\rm trk}^{\rm offline}\le 84$, and $130\le N_{\rm trk}^{\rm offline}\le 250$ are 73.0 \%, 77.3 \% and 85.3 \%, respectively. $C$ (top to bottom): $0.980\pm 0.004$, $1.031\pm 0.001$, $1.007\pm 0.001$, $1.001\pm 1\times 10^{-4}$, $1.003\pm 2\times 10^{-4}$, $1.003\pm 0.001$, $0.972\pm 0.002$, $1.028\pm 0.002$, and $1.007\pm 0.001$.} 
\vspace{2mm}
\renewcommand{\arraystretch}{1.0}
\begin{tabular}{cccccccc}
\hline
$N_{\rm trk}^{\rm offline}$
& $R_1$ (fm)
& $R_2$ (fm) 
& $\lambda_1$
& $\lambda_2$
%& $C$    
& $\alpha$
& $\beta$ (GeV$^{-2}$)
& $\chi^2$/ndf\\
\hline
  $ 0-4$
& $1.57\pm 0.15$
& $0.51\pm 0.01$
& $0.680\pm 0.034$
& $0.320\pm 0.034$
%& $0.98\pm 0.00$
& $0.062\pm 0.006$
& $0.79\pm 0.18$
&  185.4/92\\

  $ 10-12$
& $2.40\pm 0.07$
& $0.39\pm 0.02$
& $0.865\pm 0.007$
& $0.135\pm 0.007$
%& $1.03\pm 0.00$
& $0.136\pm 0.009$
& $0.99\pm 0.07$
&  137.1/92\\

  $ 31-33$
& $3.37\pm 0.07$
& $0.48\pm 0.02$
& $0.910\pm 0.004$
& $0.090\pm 0.004$
%& $1.01\pm 0.00$
& $0.048\pm 0.004$
& $1.06\pm 0.12$
&  83.3/92\\

  $ 80-84$
& $3.76\pm 0.03$
& $0.49\pm 0.01$
& $0.866\pm 0.007$
& $0.061\pm 0.001$
%& $1.00\pm 0.00$
& $0.026\pm 0.001$
& $1.53\pm 0.06$
&  81.6/92\\

  $ 105-109$
& $4.02\pm 0.06$
& $0.57\pm 0.01$
& $0.867\pm 0.149$
& $0.050\pm 0.002$
%& $1.00\pm 0.00$
& $0.020\pm 0.001$
& $1.06\pm 0.07$
&  107.0/92\\

  $ 130-250$
& $3.79\pm 0.22$
& $0.46\pm 0.09$
& $0.857\pm 0.051$
& $0.040\pm 0.011$
%& $1.00\pm 0.00$
& $0.030\pm 0.014$
& $1.54\pm 0.51$
&  77.6/92\\
\hline
\multicolumn{8}{l}{Note: When no constraint is applied for $\lambda_1$ and $\lambda_2$, we obtain the following figures:}\\
\hline
$ 0-4$
& $2.76\pm 0.30$
& $0.49\pm 0.02$
& $1.085\pm 0.083$
& $0.477\pm 0.052$
%& $0.97\pm 0.00$
& $0.112\pm 0.051$
& $1.78\pm 0.57$
&  126.6/92\\

$ 10-12$
& $2.50\pm 0.09$
& $0.37\pm 0.02$
& $0.947\pm 0.033$
& $0.168\pm 0.025$
%& $1.03\pm 0.00$
& $0.165\pm 0.028$
& $1.17\pm 0.13$
&  128.8/92\\

$ 31-33$
& $3.43\pm 0.11$
& $0.48\pm 0.02$
& $0.928\pm 0.029$
& $0.092\pm 0.004$
%& $1.01\pm 0.00$
& $0.048\pm 0.004$
& $1.06\pm 0.12$
&  83.0/92\\
\hline
\end{tabular}
\end{table}

%FIG3---------------------------------------------------------------------------

Among the six intervals shown in Fig.~\ref{fig3}, the red (solid) line and green (dashed) line appear to be exceptional. They are probably related  to the normalization factors ($0.980\pm 0.004$ and $1.031\pm 0.001$). In other words, in those regions there is very small freedom or noise which cannot be described by Eqs.~(\ref{eq2})--(\ref{eq4}).

%SECTION3-----------------------------------------------------------------------
\section{\label{sec3}Test of the two-component model for 3-pion BEC}
\begin{figure}[H]
  \centering
  \includegraphics[width=0.6\columnwidth]{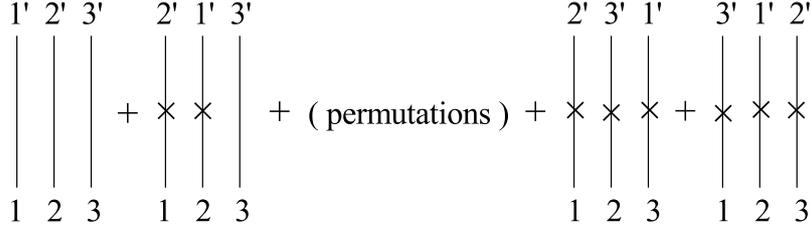}
  \caption{\label{fig4}Diagrams for the third-order BEC. The matrix indicates the exchange of identical pions.}
\end{figure}

Here, we investigate the 3-pion BEC using the two-component model. Since there is currently no information from CMS on the multiplicity distribution $P(n)$ at 13 TeV, it is challenging to determine the ratio between the contributions of the first and the second components. We use the diagrams in Fig.~\ref{fig4}.

%FIG4---------------------------------------------------------------------------

The formula that corresponds to the diagrams in Fig.~\ref{fig4}~\cite{Biyajima:1990ku,Suzuki:1999nj,Kozlov:2007dv} is expressed as
\begin{eqnarray}
F_i^{(3)}=1.0+3\lambda_iE_{\rm BE_i}+2(\lambda_iE_{\rm BE_i})^{3/2}
\label{eq5}
\end{eqnarray}
By assuming an equal weight for the first and the second components, $F_1^{(3)}$ and $F_2^{(3)}$, we obtain the following normalized expression
\begin{eqnarray}
F^{(3+:3-)}=1.0+\frac 12\left(3\lambda_1E_{\rm BE_1}+2(\lambda_1E_{\rm BE_1})^{3/2}\right)+\frac 12\left(3\lambda_2E_{\rm BE_2}+2(\lambda_2E_{\rm BE_2})^{3/2}\right),
\label{eq6}
\end{eqnarray}
where $\lambda_1$, $\lambda_2$, $R_1$, and $R_2$ are fixed by using the numerical values in Table{\ref{tab2}. Typical figures are presented in Fig.~\ref{fig5}. We could calculate the ratio if the CMS Collaboration reports the multiplicity distributions $P(n)$~\cite{CMS:2011nlc}, as this would allow us to understand the ensemble property of the BEC through the multiplicity distribution. It is worth noting that the ATLAS Collaboration has already observed the multiplicity distributions $P(n)$~\cite{ATLAS:2016zba} and BEC~\cite{ATLAS:2022wvk} to be considered in~\cite{Biyajima:2023aaa}.

%FIG5---------------------------------------------------------------------------
\begin{figure}[H]
  \centering
  \includegraphics[width=0.45\columnwidth]{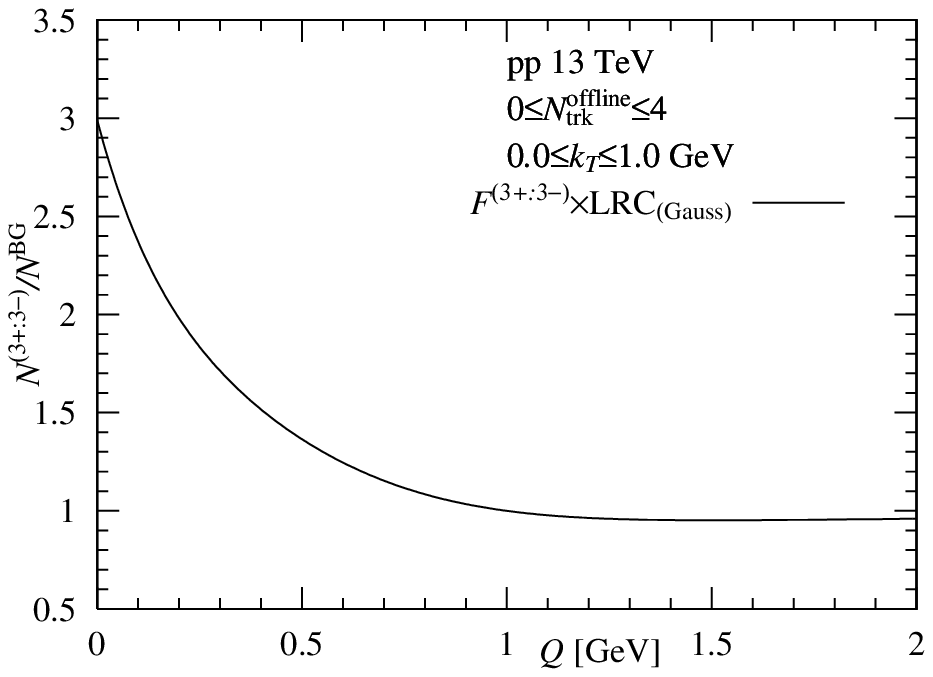}
  \includegraphics[width=0.45\columnwidth]{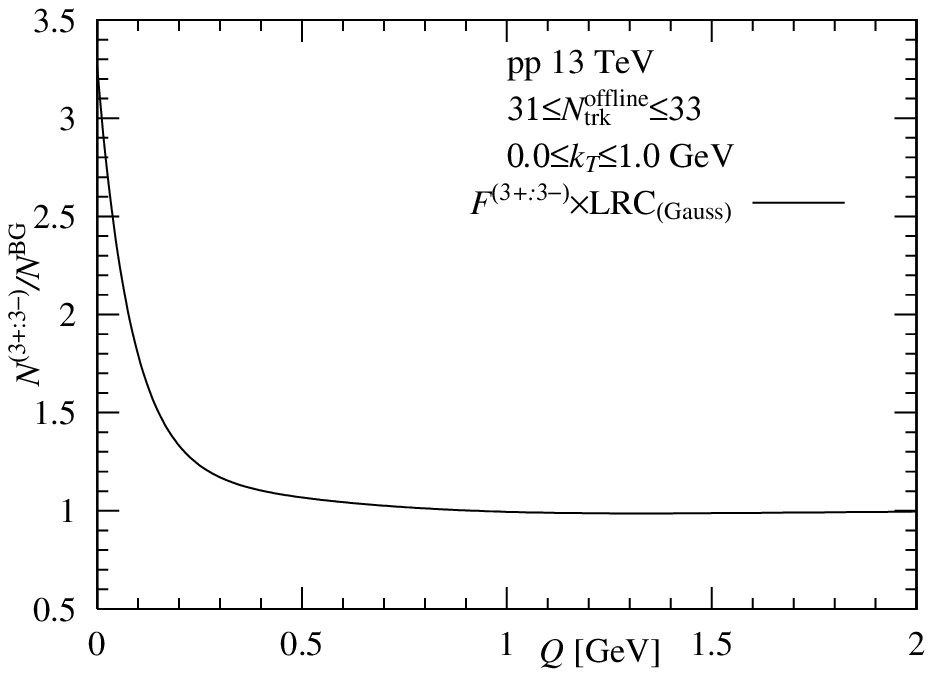}
  \includegraphics[width=0.45\columnwidth]{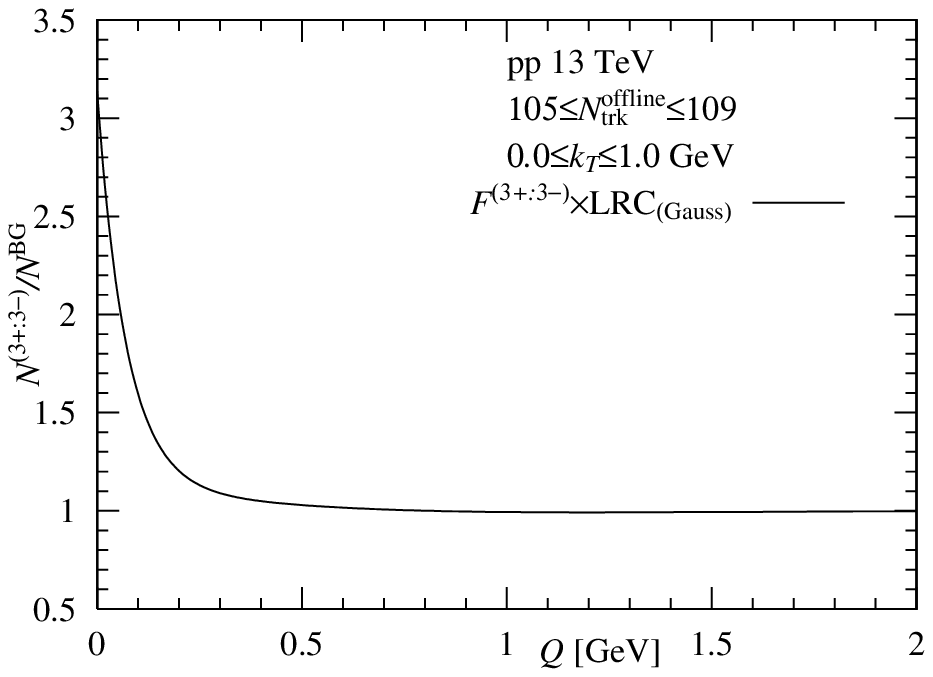}
  \caption{\label{fig5}Prediction of upper limit of the 3$\pi$ BEC in pp collisions at 13 TeV by means of Eq.~(\ref{eq6}) with Eqs.~(\ref{eq2})-(\ref{eq4}). $N^{\rm (BG)}$ means $N_{\rm MC}^{(3+:3-)}$, because of no-BEC in the Monte Carlo events.}
\end{figure}

In the near future, we may be able to further test the two-component model when the CMS Collaboration analyzes the 3-$\pi$ BEC. If we observe the same extensions as in Fig.~\ref{fig2}, we could conclude that the two-component model is a viable approach.

%SECTION4-----------------------------------------------------------------------
\section{\label{sec4}Concluding remarks}
\begin{description}
  \item[C1)] Our analysis of CMS BEC at 13 TeV using the $\tau$-model with Eq.~(\ref{eq1}) confirms the applicability of this model. This is evidenced by the values of $\chi^2$ in Table~\ref{tab1}.

  \item[C2)] As portrayed in Table~\ref{tab1}, the interaction ranges $r$ in the L\'evy-type expression $e^{-(Qr)^{\alpha_{\tau}}}$ increase as the range of the interval $N_{\rm trk}^{\rm offline}$ increases. However, it appears that the interaction ranges from 30 to 50 fm are large in $pp$ collisions at 13 TeV.

  \item[C3)] To gain a better understanding of the results obtained from the $\tau$-model, we have analyzed the BEC using the $\tau$-model with Eq.~(\ref{eq4}). This has led to improved estimations, as shown in Appendix~\ref{secA}. 

  \item[C4)] We look forward to future analyses by the CMS Collaboration
 of the multiplicity distributions and the third-order BEC at 13 TeV \end{description}

Hereafter, we summarize the results of the two-component model using Eqs.~(\ref{eq2})--(\ref{eq4}).      

\begin{description}
  \item[C5)] In Table~\ref{tab1A}, we mentioned how to propose LRC$_{(\rm Gauss)}$, i.e., Eq.~(\ref{eq4}). To investigate the remarks mentioned in C2) above using the two-component model, we utilized Eqs.~(\ref{eq2})--(\ref{eq4}). Our results are presented in Table~\ref{tab2}. The large extensions are approximately 4 fm, and they appear to be reasonable.

  \item[C6)] Furthermore, to test the availability of the two-component model, we calculated the 3-pion BEC by making use of the estimated values and diagrams presented in Fig.~\ref{fig4}. Interestingly, as $N_{\rm trk}^{\rm offline}$ increases, the 3-pion BEC rapidly decreases, due to the changes in the extension $R_1$ (1 fm to 4 fm). Moreover, the intercepts at $Q=0.0$ GeV are about 3.0, providing the equal weight.

  \item[C7)] To investigate the role of the LRC$_{(\rm Gauss)}$, i.e., Eq.~(\ref{eq4}), we reanalyzed the BEC at 0.9 and 7 TeV, with the results presented in Appendix~\ref{secB}. The estimated $\chi^2$ values became smaller than that of LRC$_{(\rm linear)}$~\cite{Biyajima:2019wcb}. We also analyzed the data with the constraint, $0.1\le k_T\le 0.3$ GeV and $2\le N_{\rm ch}\le 9$. We observe that $R_1$'s increase and $R_2$'s are almost constant.

  \item[C8)] As portrayed in Table~\ref{tab2}, the BEC in the intervals $0\le N_{\rm trk}^{\rm offline}\le 4$ and $10\le N_{\rm trk}^{\rm offline}\le 12$ cannot be analyzed with better $\chi^2$ values. A more complicated model may be necessary.

  \item[C9)]From Table~\ref{tab2}, we can observe behaviors of $R_1$ and $R_2$'s in Fig.~\ref{fig6}. The larger extension $R_1$'s seem to be saturated at larger $N_{\rm trk}^{\rm offline}$. To confirm that, of course, more data are needed. Compare Fig.~\ref{figB3} and Table~\ref{tabB2} in Appendix B, where data at 7 TeV with the constraints $0.1\le k_T\le 0.3$ GeV and and $2\le N_{\rm ch}\le 9$ are analyzed. See also discussions on two kinds of extensions mentioned in Ref.~\cite{Khoze:2016hns}.
\end{description}
%
%FIG6---------------------------------------------------------------------------
\begin{figure}[H]
  \centering
  \includegraphics[width=0.6\columnwidth]{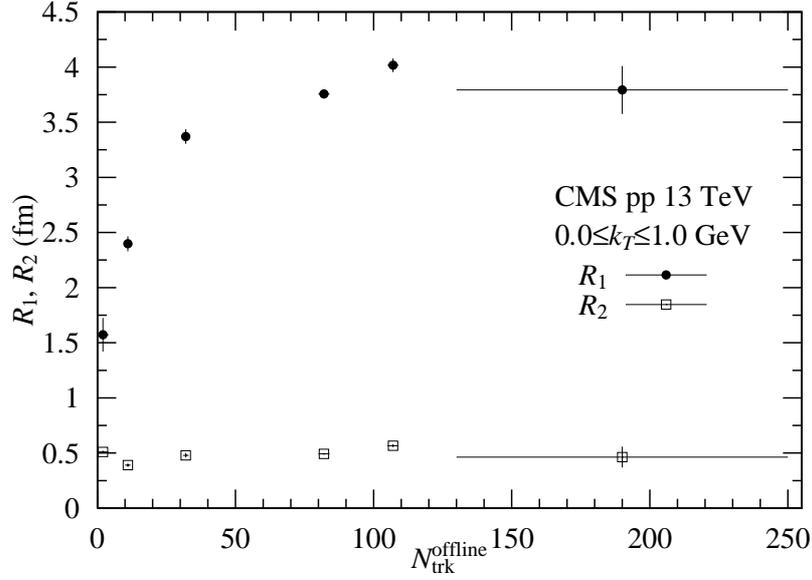}
  \caption{\label{fig6}Behaviors of $R_1$ and $R_2$'s estimated in Table~\ref{tab2}. The smaller extension $R_2$'s are almost the constant.}
\end{figure}

\noindent
{\it Acknowledgments.} One of the authors (M.B.) would like to thank his colleagues at the Department of Physics, Shinshu University.

%APPENDIX A---------------------------------------------------------------------

\appendix

\section{\label{secA}Analysis of BEC at 13 TeV using the $\tau$-model with Eq.~(\ref{eq4})}
We are interested in the influence of Eq.~(\ref{eq4}) on the $\tau$-model. To investigate this, we reanalyzed the BEC using the following formula
\begin{eqnarray}
F_{\tau\mathchar`-{\rm Gauss}} = [1+\lambda\cos( (r_0 Q)^2+\tan(\alpha_{\tau}\pi/4)(Qr)^{\alpha_{\tau}})e^{-(Qr)^{\alpha_{\tau}}}]\cdot {\rm LRC_{(Gauss)}}
\label{eqA1}
\end{eqnarray}
Our findings are presented in Fig.~\ref{figA1} and Table~\ref{tabA1}. It can be seen that the interaction range $r$ values are smaller than 10 fm. 
%
%FIGA1---------------------------------------------------------------------------
\begin{figure}[H]
  \centering
  \includegraphics[width=0.48\columnwidth]{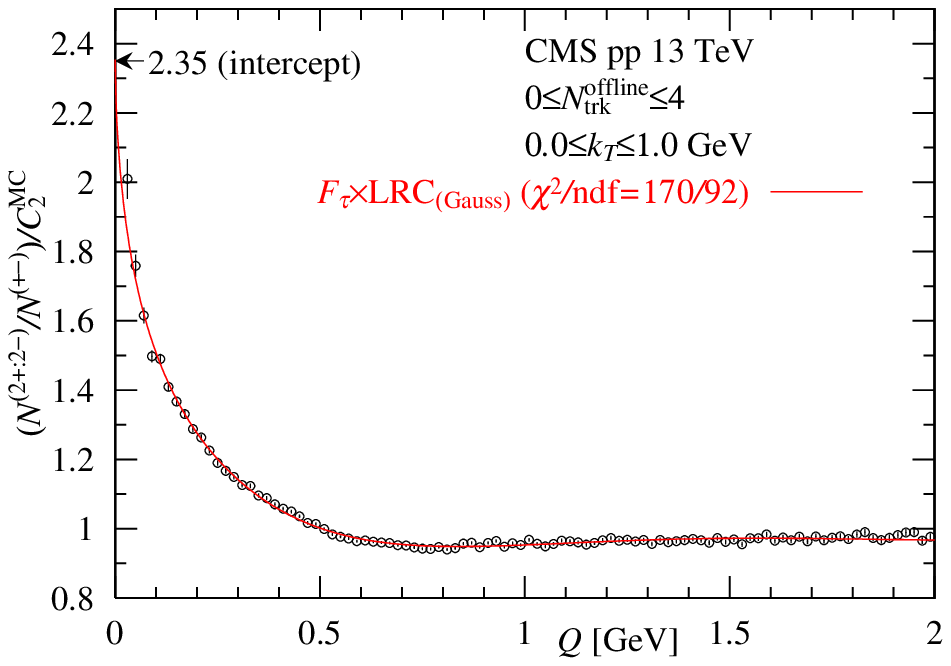}
  \includegraphics[width=0.48\columnwidth]{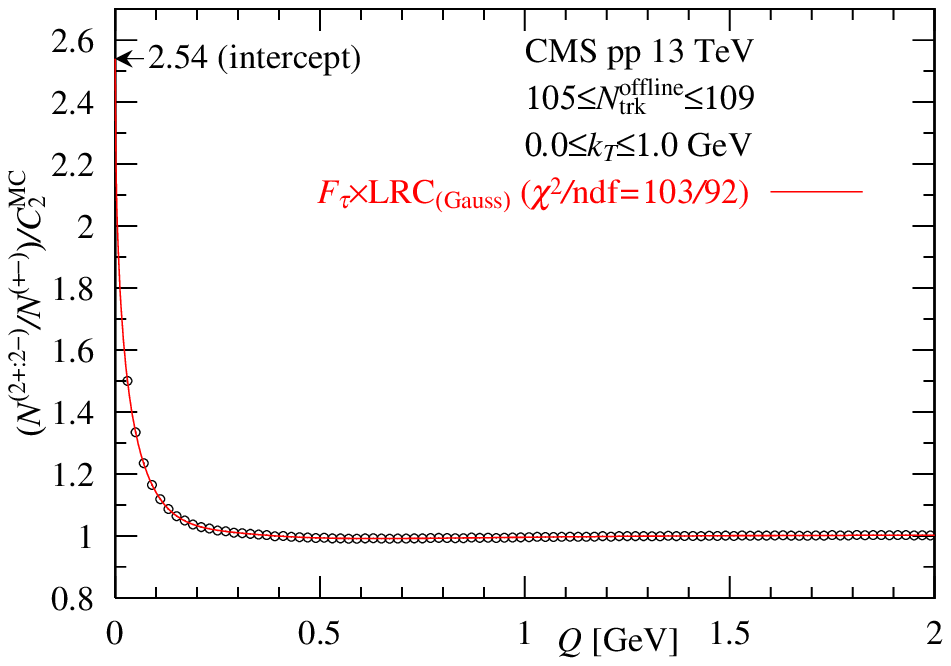}
  \caption{\label{figA1}Fit to the BEC measurements by CMS in pp collisions at 13 TeV by Eq.~(\ref{eqA1}) with Eq.~(\ref{eq4}).}
\end{figure}

%TABA1---------------------------------------------------------------------------
\begin{table}[H]
\centering
\caption{\label{tabA1}Fit parameters of the CMS measurements of BEC in pp collisions at 13 TeV ($0.0\le k_T\le 1.0$ GeV) using the $\tau$-model with Eq.~(\ref{eq4}).} 
\vspace{2mm}
\renewcommand{\arraystretch}{1.0}
\begin{tabular}{cccccccc}
\hline
$N_{\rm trk}^{\rm offline}$
& $r_0$ (fm)
& $r$ (fm) 
& $\lambda$
%& $C$    
& $\alpha_{\tau}$
& $\alpha$
& $\beta$
& $\chi^2$/ndf\\
\hline
  $ 0-4$
& $ 0.22\pm 0.02$
& $ 2.12\pm 0.52$
& $ 1.02\pm 0.12$
%& $ 0.97\pm 0.00$
& $ 0.595\pm 0.047$
& $\!\!\!\!-0.169\pm 0.017$
& $ 5.60\pm 0.49$
&  169.6/92\\

  $ 10-12$
& $ 0.25\pm 0.01$
& $ 9.89\pm 1.69$
& $ 2.49\pm 0.25$
%& $ 1.08\pm 0.06$
& $ 0.417\pm 0.014$
& $ 0.099\pm 0.060$
& $ 0.17\pm 0.13$
&  138.7/92\\

  $ 31-33$
& $ 0.16\pm 0.01$
& $ 3.27\pm 0.45$
& $ 1.76\pm 0.10$
%& $ 1.00\pm 0.00$
& $ 0.566\pm 0.022$
& $ 0.177\pm 0.024$
& $19.86\pm 1.02$
&   85.9/92\\

  $ 80-84$
& $ 0.00$
& $ 3.76\pm 0.06$
& $ 1.65\pm 0.02$
%& $ 1.00\pm 0.00$
& $ 0.566\pm 0.002$
& $ 0.159\pm 0.003$
& $23.83\pm 0.31$
&  166.2/92\\

  $ 105-109$
& $ 0.13\pm 0.01$
& $ 5.61\pm 0.54$
& $ 1.84\pm 0.08$
%& $ 1.00\pm 0.00$
& $ 0.517\pm 0.012$
& $ 0.119\pm 0.010$
& $25.48\pm 0.73$
&  103.1/92\\

  $ 130-250$
& $ 0.18\pm 0.02$
& $ 9.02\pm 3.43$
& $ 2.18\pm 0.42$
%& $ 1.00\pm 0.00$
& $ 0.468\pm 0.037$
& $ 0.076\pm 0.024$
& $21.63\pm 3.39$
&  78.5/92\\
\hline
\end{tabular}
\end{table}

%FIGA2---------------------------------------------------------------------------
\begin{figure}[H]
  \centering
  \includegraphics[width=0.48\columnwidth]{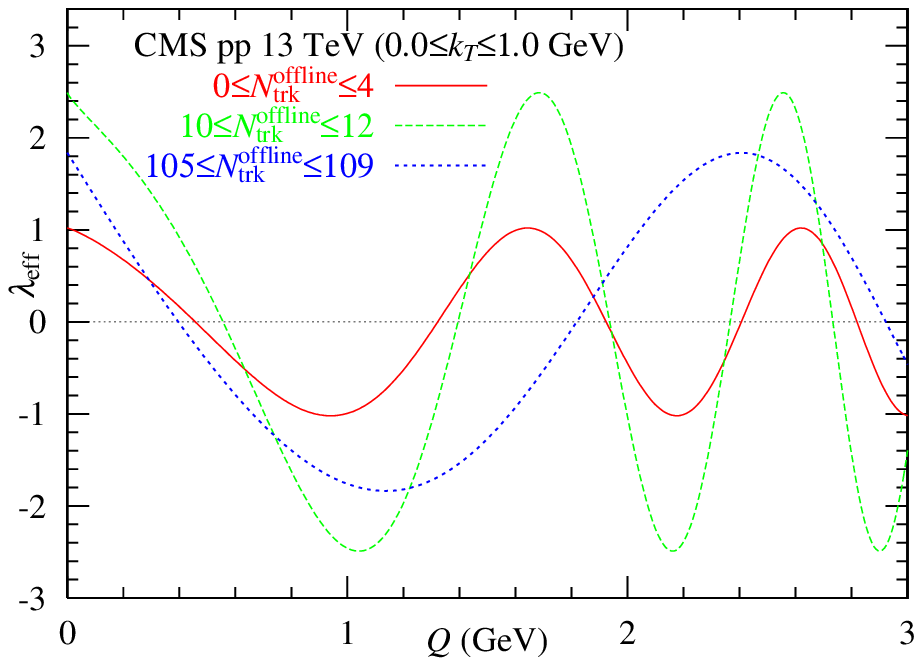}
  \includegraphics[width=0.48\columnwidth]{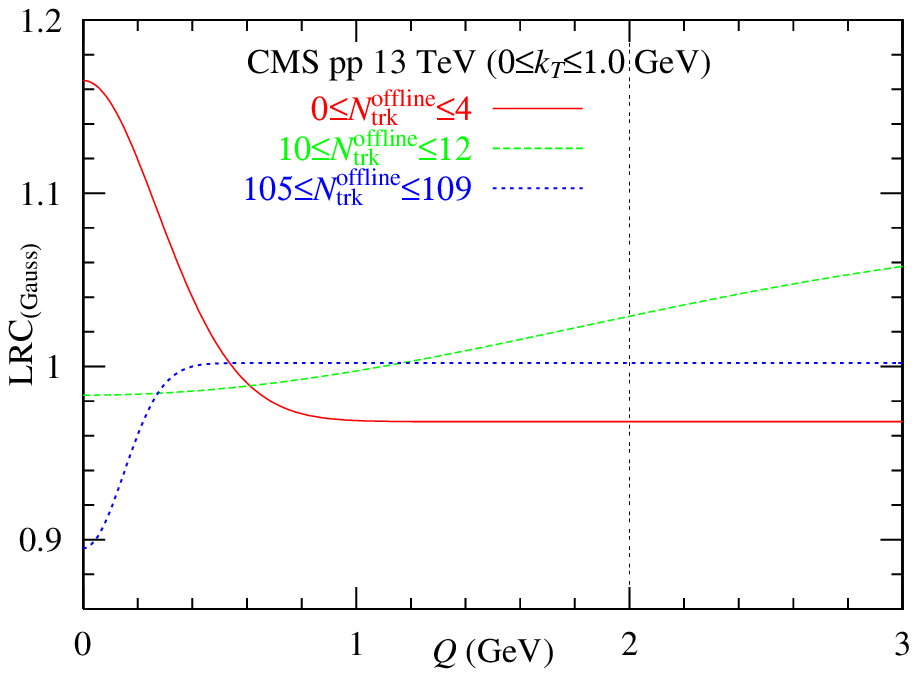}
  \caption{\label{figA2}$\lambda_{\rm eff}$'s and LRCs of BEC measurements by CMS in pp collisions at 13 TeV by Eq.~(\ref{eqA1}). The vertical line at $Q=2.0$ GeV represents the effective range of the LRC ($0\le Q\le 2$ GeV).}
\end{figure}

As illustrated in Fig.~\ref{figA2}, three LRC's appear to be various.  Therein the behavior of LRC for $0\le N_{\rm trk}^{\rm offline}\le 4$ is related to the negative $\alpha$. For the sake of reference, we demonstrate the effective degree of coherence in the $\tau$-model, 
$$
\lambda_{\rm eff}=\lambda\cos( (r_0 Q)^2+\tan(\alpha_{\tau}\pi/4)(Qr)^{\alpha_{\tau}})
$$
in Fig.~\ref{figA2}. By making use of $\lambda_{\rm eff}$'s and LRCs, we can estimate the intercepts at $Q=0.0$ GeV, which are shown in Fig.~\ref{figA1}. 

%APPENDIX B---------------------------------------------------------------------
\section{\label{secB}Reanalysis of CMS BEC at 0.9 and  7 TeV~\cite{CMS:2011nlc} by LRC, expressed by Eq.~(\ref{eq4})}
We examined the changes in the values of $\chi^2$ when LRC$_{(\rm linear)}$ was replaced with Eq.~(\ref{eq4}) in the reanalysis of BEC at 0.9 and 7 TeV~\cite{CMS:2011nlc}. Our new results obtained using Eq.~(\ref{eq4}) are presented in Fig.~\ref{figB1} and Table~\ref{tabB1} and compared with those obtained elsewhere~\cite{Biyajima:2019wcb}, where the linear form for the ${\rm LRC}=C(1+\delta Q)$ was used. These results are also shown in Table~\ref{tabB1}. We show the LRCs in Fig.~\ref{figB2}.

%FIGB1---------------------------------------------------------------------------
\begin{figure}[H]
  \centering
  \includegraphics[width=0.48\columnwidth]{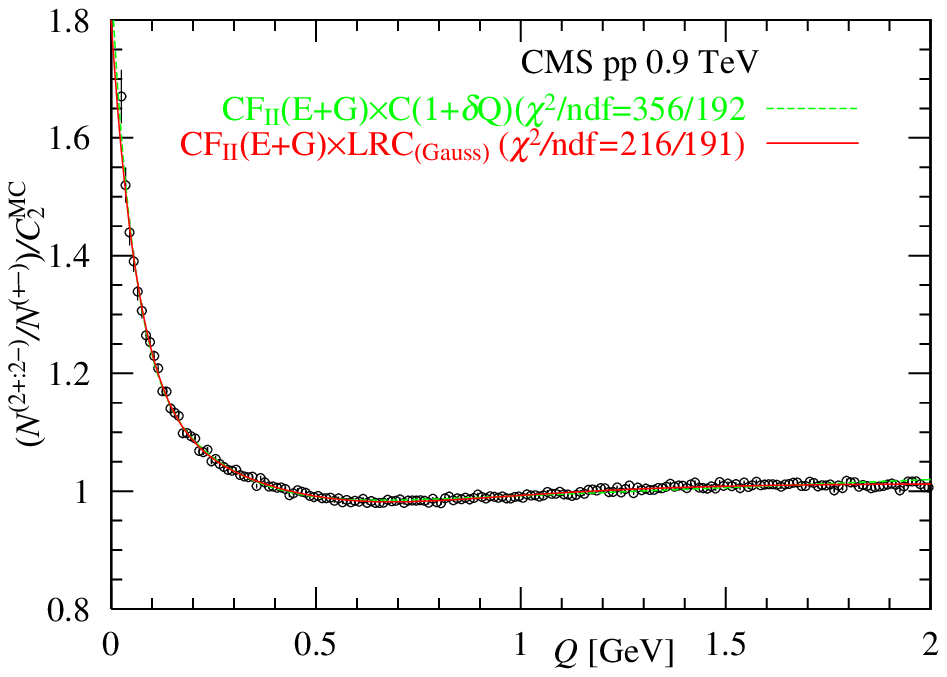}
  \includegraphics[width=0.48\columnwidth]{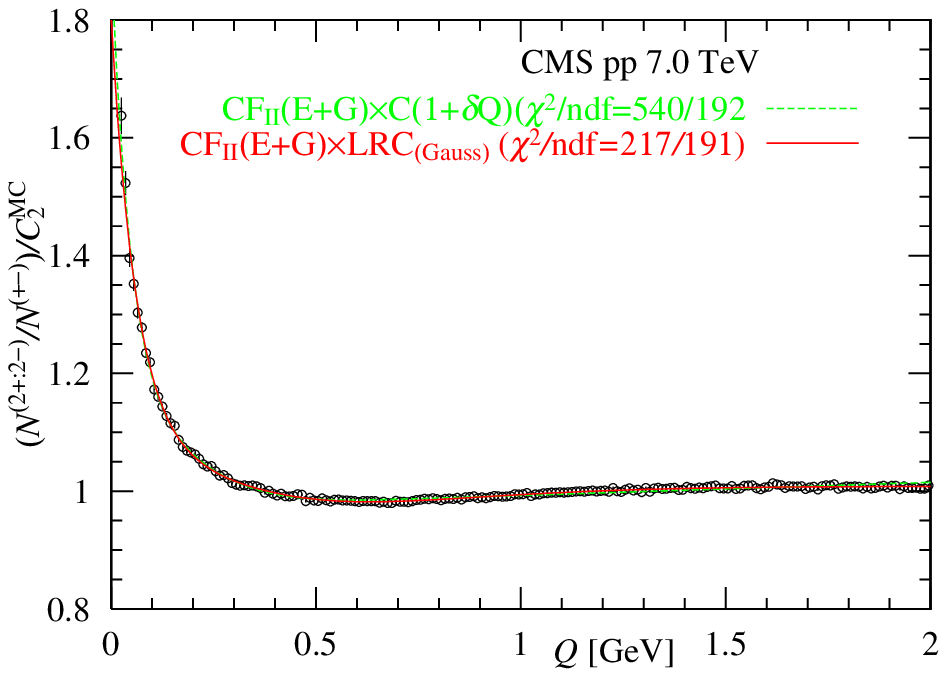}
  \caption{\label{figB1}Fit to the CMS BEC measurements in pp collisions at 0.9 and 7.0 TeV by Eqs.~(\ref{eq2})--(\ref{eq4}).}
\end{figure}

%TABB1---------------------------------------------------------------------------
\begin{table}[H]
\centering
\caption{\label{tabB1}Fit parameters of the CMS BEC measurements in pp collisions at 0.9 and 7.0 TeV by Eqs.~(\ref{eq2})--(\ref{eq4}).} 
\vspace{2mm}
\renewcommand{\arraystretch}{1.0}
\begin{tabular}{ccccccc}
\hline
& $R_1$ (fm)
& $R_2$ (fm) 
& $\lambda_1$
& $\lambda_2$
%& $C$    
%& $\alpha$
%& $\beta$ (GeV$^{-2}$)
& $\!\!\!\delta\,({\rm GeV}^{-1})$ or $(\alpha,\ \beta\,({\rm GeV}^{-2}))\!\!\!$
& $\chi^2$/ndf\\
\hline
$\sqrt s=0.9$ TeV$\!\!\!$\\
\hline
LRC$_{\rm (linear)}$~\cite{Biyajima:2019wcb}
& $1.42\pm 0.03$
& $3.37\pm 0.42$
& $0.53\pm 0.02$
& $0.23\pm 0.04$
%& $0.962\pm 0.001$
& $0.030\pm 0.001$
%& ---
& 417/192\\
Eq.~(\ref{eq4})
& $0.82\pm 0.10$
& $2.27\pm 0.13$
& $0.41\pm 0.01$
& $0.30\pm 0.02$
%& $1.012\pm 0.001$
& $(0.10\pm 0.02,\ 
 1.38\pm 0.10)$
& 234/191\\
\hline
$\sqrt s= 7$ TeV\\
\hline
LRC$_{\rm (linear)}$~\cite{Biyajima:2019wcb}
& $3.88\pm 0.18$
& $0.71\pm 0.01$
& $0.84\pm 0.03$
& $0.12\pm 0.01$
%& $0.97\pm 0.00$
& $0.023\pm 0.001$
%& ---
& 540/192\\
Eq.~(\ref{eq4})
& $3.13\pm 0.13$
& $0.51\pm 0.02$
& $0.80\pm 0.03$
& $0.10\pm 0.01$
%& $1.01\pm 0.00$
& $(0.06\pm 0.01,\ 
 1.46\pm 0.11)$
& 217/191\\
\hline
\end{tabular}
\end{table}

%FIGB2---------------------------------------------------------------------------
\begin{figure}[H]
  \centering
  \includegraphics[width=0.6\columnwidth]{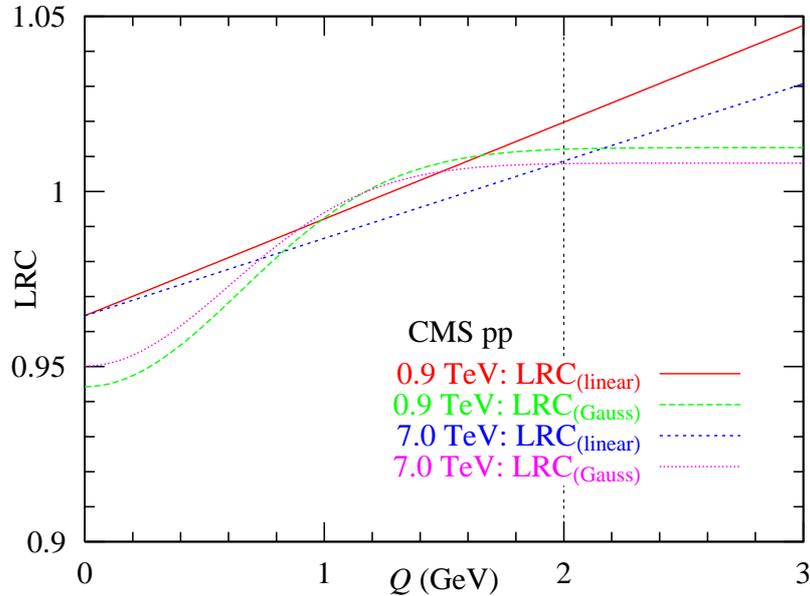}
  \caption{\label{figB2}LRCs by LRC$_{\rm (linear)}$~\cite{Biyajima:2019wcb} and LRC$_{(\rm Gauss)}$ (Eq.~(\ref{eq4})) of CMS BEC measurements in pp collisions at 0.9 and 7.0 TeV are presented. The vertical line at $Q=2.0$ GeV represents the effective range of the LRC ($0\le Q\le 2$ GeV).}
\end{figure}

It can be said that the Gaussian distribution of the LRC in the two-component model is better than that of the linear form, because the LRC$_{(\rm Gauss)}$ converges to 1.0 in the region $Q\ge 2.0$ GeV. The reason is as follows: The emitting source functions and/or the LRC's in the Euclidean space ($Q' = \sqrt{({\bf p}_1-{\bf p}_2)^2+(E_1-E_2)^2}$ and $\xi' = \sqrt{({\bf r}_1-{\bf r}_2)^2+(t_1-t_2)^2}$) are calculated as 
\begin{eqnarray}
F_{\rm source}(\xi',\,R) = \frac 1{(2\pi)^2\xi'} \int_0^{\infty} Q'^2\, E_{\rm BE}(Q',\,R)\, J_1(Q'\xi')dQ'
\label{eqA2}
\end{eqnarray}
where $J_1(Q\xi)$ is the Bessel function. For the LRC, we should replace $E_{\rm BE}$ with (LRC$-1.0$) and $R$ with $\beta$ in Eq.~(\ref{eqA2}), respectively. In other words, the $({\rm LRC_{(Gauss)}-1.0})=\sum_{k=1}^{\infty} (-\alpha e^{-\beta Q^2})^k$ is preferable to the $({\rm LRC_{(linear)}-1.0)}=\delta Q$, because the former converges, as $Q$ is large. Finally, we should adopt the inverse Wick rotation for $\xi'$~\cite{Shimoda:1992gb,Mizoguchi:2021slz}; $\xi = \sqrt{({\bf r}_1-{\bf r}_2)^2-(t_1-t_2)^2}$.

Finally we analyzed data on BEC at 7 TeV with $0.1\le k_T\le 0.3$ GeV and $2\le N_{\rm ch}\le 9$ in Ref.~\cite{CMS:2011nlc} by means of Eqs.~(\ref{eq2})-(\ref{eq4}). The smaller extensions are almost constant. This fact is similar to Fig.~\ref{fig6}. From estimated parameters with the constraint $2\le N_{\rm ch}\le 9$ (fixed), we know that both $R_1$ and $R_2$'s are approximately constant.

%FIGB3---------------------------------------------------------------------------
\begin{figure}[H]
  \centering
  \includegraphics[width=0.48\columnwidth]{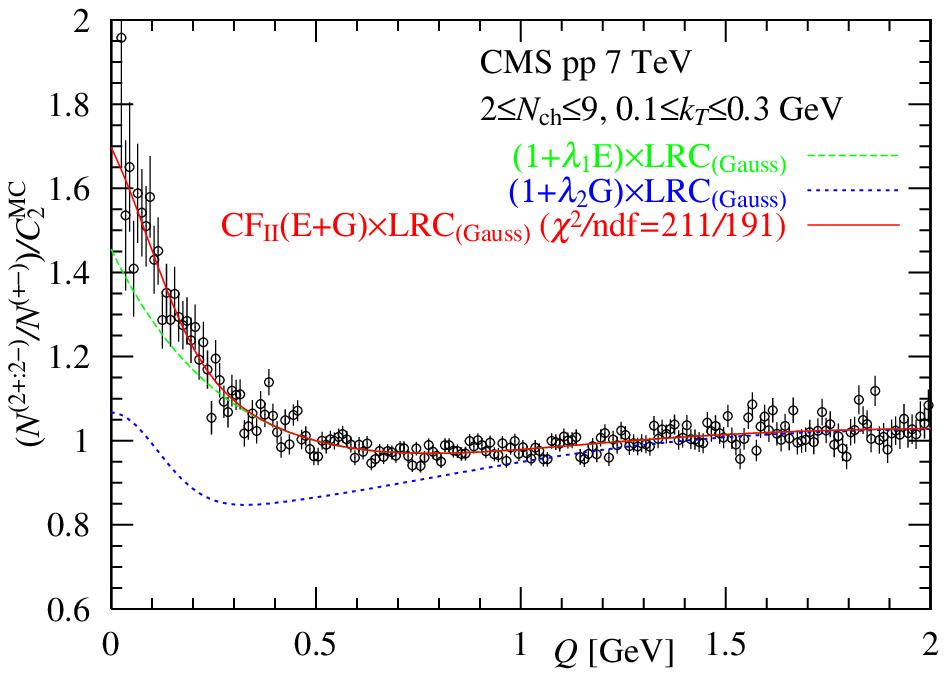}
  \includegraphics[width=0.48\columnwidth]{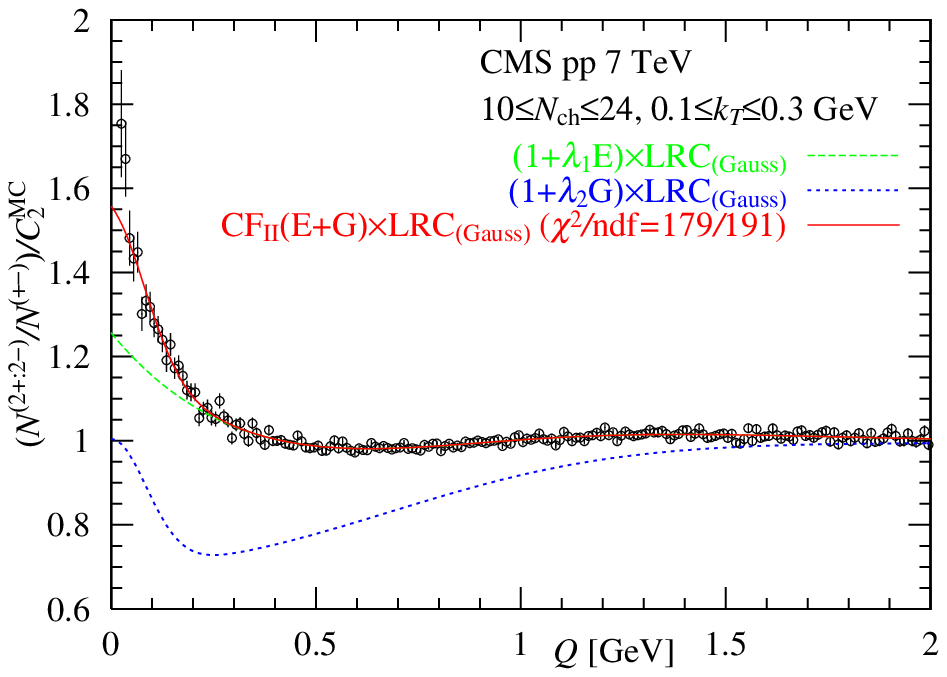}\\
  \includegraphics[width=0.48\columnwidth]{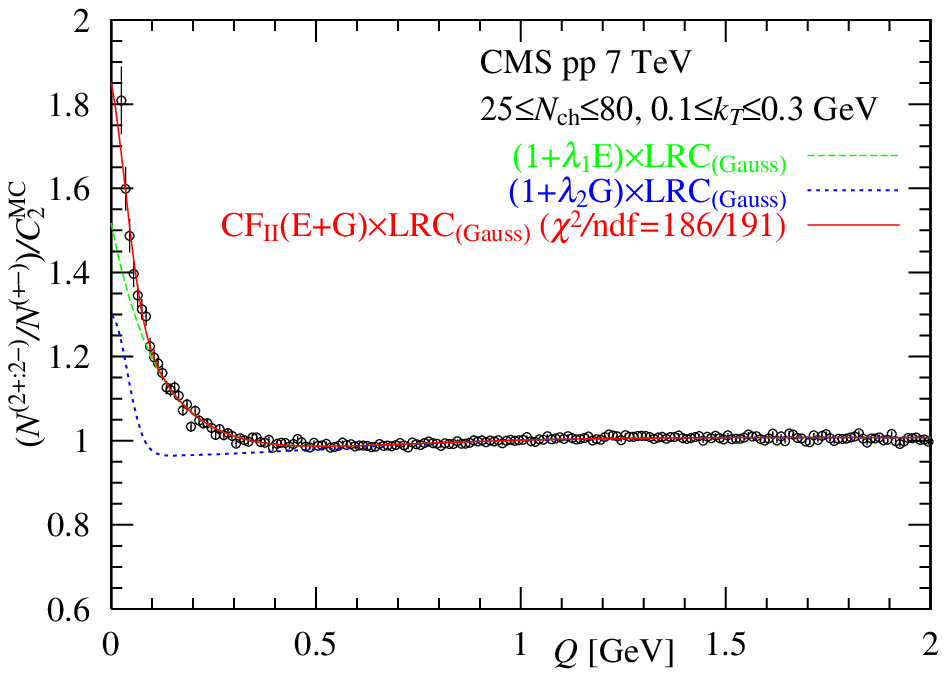}
  \caption{\label{figB3}Fit to the BEC measurements by CMS in pp collisions at 7 TeV with $0.1\le k_T\le 0.3$ GeV by Eqs.~(\ref{eq2})--(\ref{eq4}).}
\end{figure}

%TABB2---------------------------------------------------------------------------
\begin{table}[H]
\centering
\caption{\label{tabB2}Fit parameters of the CMS BEC measurements in pp collisions at 7.0 TeV with $0.1\le k_T\le 0.3$ GeV and $2\le N_{\rm ch}\le 9$ by Eqs.~(\ref{eq2})--(\ref{eq4}) with $0\le \lambda_1\le 1$ and $\lambda_2=1-\lambda_1$. $N_{\rm ch}$ means the charged particle multiplicity. $C$ (top to bottom): $1.037\pm 0.011$, $1.013\pm 0.001$, $1.008\pm 0.001$, $1.010\pm 0.008$, and $1.037\pm 0.011$.} 
\vspace{2mm}
\renewcommand{\arraystretch}{1.0}
\begin{tabular}{cccccccc}

\hline
& $R_1$(fm)
& $R_2$(fm) 
& $\lambda_1$
& $\lambda_2$
%& $C$
& $\alpha$
& $\beta$ (GeV$^{-2}$)
& $\chi^2$/ndf\\

\hline
\multicolumn{2}{l}{$0.1\le k_{T}\le 0.3$ (fixed)}\\
\hline

$2\le N_{\rm ch}\le 9$
& $1.16\pm 0.31$
& $0.36\pm 0.19$
& $0.95\pm 0.10$
& $0.05\pm 0.10$
%& $1.037\pm 0.011$
& $0.14\pm 0.04$
& $0.80\pm 0.28$
& 212/192\\

$10\le N_{\rm ch}\le 24$
& $2.05\pm 0.25$
& $0.46\pm 0.04$
& $0.95\pm 0.10$
& $0.05\pm 0.10$
%& $1.013\pm 0.001$
& $0.14\pm 0.03$
& $2.42\pm 0.24$
& 188/192\\

$25\le N_{\rm ch}\le 80$
& $2.77\pm 0.14$
& $0.47\pm 0.04$
& $0.91\pm 0.02$
& $0.09\pm 0.02$
%& $1.008\pm 0.001$
& $0.08\pm 0.02$
& $2.47\pm 0.26$
& 176/192\\

\hline
\multicolumn{2}{l}{$2\le N_{\rm ch}\le 9$ (fixed)}\\
\hline
$0.1\le k_T\le 0.3$
& $1.16\pm 0.31$
& $0.36\pm 0.19$
& $0.95\pm 0.10$
& $0.05\pm 0.10$
%& $1.037\pm 0.011$
& $0.14\pm 0.04$
& $0.80\pm 0.28$
& 212/192\\

$0.3\le k_T\le 0.5$
& $1.47\pm 0.20$
& $0.29\pm 0.05$
& $0.86\pm 0.03$
& $0.14\pm 0.03$
%& $1.010\pm 0.008$
& $0.15\pm 0.04$
& $0.99\pm 0.23$
& 198/192\\

$0.5\le k_T\le 1.0$
& $0.97\pm 0.41$
& $0.25\pm 0.07$
& $0.71\pm 0.10$
& $0.29\pm 0.10$
%& $1.037\pm 0.011$
& $0.43\pm 0.08$
& $1.16\pm 0.32$
& 177/192\\

\hline
\end{tabular}
\end{table}

%BIBLIOGRAPHY-------------------------------------------------------------------

\end{document}